\documentclass[pra,twocolumn,superscriptaddress,shownopacs,amssymb]{revtex4-2}
\usepackage{graphicx,psfrag,amsmath,amssymb}

\usepackage[usenames, dvipsnames]{color}

\usepackage{hyperref} 
\hypersetup{
    colorlinks=true,
    linkcolor=Blue,
    citecolor=Blue,
    filecolor=Blue,      
    urlcolor=Blue,
    pdftitle={BEC two-band},
    }

\begin{document}

\title{Quantum-geometric contribution to the Bogoliubov modes \\
in a two-band Bose-Einstein condensate}

\author{M. Iskin}
\affiliation{
Department of Physics, Ko\c{c} University, Rumelifeneri Yolu, 
34450 Sar\i yer, Istanbul, Turkey
}

\date{\today}

\begin{abstract}

We consider a weakly-interacting Bose-Einstein condensate (BEC) that is loaded 
into an optical lattice with a two-point basis, and described by a two-band 
Bose-Hubbard model with generic one-body and two-body terms. 
By first projecting the system onto the lower Bloch band and then applying the 
Bogoliubov approximation to the resultant Hamiltonian, we show that the inverse
effective-mass tensor of the superfluid carriers in the Bogoliubov spectrum has 
two physically distinct contributions. In addition to the usual inverse band-mass 
tensor that is originating from the intraband processes within the lower Bloch band, 
there is also a quantum-geometric contribution that is induced by the two-body 
interactions through the interband processes. We also discuss the conditions 
under which the latter contribution is expressed in terms of the quantum-metric 
tensor of the Bloch states, i.e., the natural Fubini-Study metric on the Bloch sphere.

\end{abstract}

\maketitle

\section{Introduction}
\label{sec:intro}

Recent theoretical efforts have firmly established that the quantum geometry 
of the underlying Bloch states lies at the heart of some multiband Fermi 
superfluids, e.g., Refs.~\cite{peotta15, liang17, huhtinen22, herzog22, chan22b} 
and the references therein. 
This is because, by controlling the effective-mass tensor of the superfluid 
carriers through the interband transitions~\cite{iskin18b, huhtinen22, herzog22}, 
the quantum geometry can affect those superfluid properties that has explicit 
dependence on the carrier mass, e.g., superfluid weight/density, 
Berezinskii-Kosterlitz-Thouless transition temperature, and low-energy collective 
excitations~\cite{peotta15, liang17, huhtinen22, herzog22, chan22b, iskin18b, iskin20b}. 
While these predictions are based 
heavily on the multiband extension of the BCS formulation of superconductivity, 
their physical origins root deep into the two-body problem which is exactly 
tractable~\cite{iskin21, iskin22}.
For instance a pair of particles can still acquire a finite effective mass from 
the quantum geometry in case when its unpaired constituents are completely localized 
and immobile, e.g., due to their diverging band mass in a flat 
band~\cite{torma18, iskin21, iskin22}. 
Thus quantum-geometric interband mechanism resolves how superfluidity of Cooper 
pairs can prevail in a flat band~\cite{peotta15}.

Despite all these progress with Fermi superfluids, the defining effects of 
quantum geometry on multiband Bose superfluids is still in its infancy. 
For instance, in the case of a weakly-interacting BEC in a flat Bloch band, 
there are several multiband Bogoliubov analyses revealing that the quantum 
geometry dictates the speed of sound, quantum depletion, density-density 
correlations, and superfluid weight in fundamentally different 
ways~\cite{julku21a, julku21b, julku22}. 
In addition there is a similar analysis for the spin-orbit-coupled Bose 
superfluids highlighting that the quantum geometry of the helicity states 
in such a two-band continuum model plays an analogous role~\cite{subasi22},
which is in accordance with the recent works on the spin-orbit-coupled 
Fermi superfluids~\cite{iskin18a, iskin18b}.

Motivated by these recent works, and given that a crystal structure with 
a two-point basis is the minimal model to study quantum geometry of the 
Bloch states, here we consider a two-band Bose-Hubbard model with generic 
single-particle spectrum and two-body interactions. 
By deriving the Bogoliubov spectrum for the low-energy quasiparticle 
excitations, we show that the interband processes 
that are induced by the two-body interactions give rise to a 
quantum-geometric contribution and dress the effective-mass tensor 
of the superfluid carriers. 
In the particular case when BEC occurs uniformly within a unit cell, we also 
relate the geometric contribution to the quantum-metric tensor of the 
Bloch states. Similar to the Bogoliubov spectrum and superfluid 
weight/density, it is conceivable that all of the superfluid properties that 
depend on the carrier mass also have analogous quantum-geometric contributions. 
In fact the energetic stability of a weakly-interacting Bose superfluid relies 
solely on these contributions when the BEC occurs in a flat Bloch 
band~\cite{julku21a, julku21b, julku22}.

The remaining text is organized as follows. In Sec.~\ref{sec:BHm} we introduce 
the two-band Bose-Hubbard Hamiltonian in momentum space, and project it to the 
lower Bloch band. Then in Sec.~\ref{sec:Bt} we apply the Bogoliubov approximation 
to the projected Hamiltonian, and extract the low-energy Bogoliubov modes. 
There we also relate the effective-mass tensor of the superfluid carriers
to the quantum-metric tensor of the Bloch states, and compare their relation 
with that of the Fermi superfluids. In Sec.~\ref{sec:conc} we end the paper 
with a brief summary of our findings and an outlook. Benchmark with the 
extended Bose-Hubbard model is briefly discussed in App. A, and some example 
models with non-trivial geometry are presented in App. B.

\section{Bose-Hubbard model}
\label{sec:BHm}

In the presence of multiple sublattices, a generic Hamiltonian for the 
Bose-Hubbard model can be written as
\begin{align}
\label{eqn:BHHam}
\mathcal{H} &= -\sum_{Si; S'i'} t_{Si; S'i'} c_{S i}^\dagger c_{S' i'}
+ \frac{1}{2} \sum_{S i} U_S c_{S i}^\dagger c_{S i}^\dagger c_{S i} c_{S i} 
\nonumber \\
& + \sum_{\langle S i, S'i' \rangle} V_{S i, S'i'} c_{S i}^\dagger c_{S' i'}^\dagger c_{S' i'} c_{S i} 
- \mu \sum_{Si} c_{S i}^\dagger c_{S i},
\end{align}
where the operator $c_{S i}^\dagger$ ($c_{S i}$) creates (annihilates) a particle 
at the sublattice site $S$ in the unit cell $i$, the hopping parameter $t_{Si; S'i'}$ 
characterizes the tunneling between any pair of sites, the two-body terms $U_S$ 
and $V_{S i, S'i'}$ describe, respectively, the onsite and offsite density-density 
interactions, and $\mu$ is the chemical potential. 
It is only the onsite interaction that is considered in the previous 
works~\cite{julku21a, julku21b, julku22}. The number of $S$ sites 
determines the number of Bloch bands in the single-particle spectrum, and a 
lattice with a two-point basis (i.e., $S \in \{A, B\}$) is the minimal model to 
study the quantum geometry of underlying Bloch states. For the sake of clarity 
and for its simplicity, here we restrict our analysis to a two-band Bose-Hubbard 
model with generic single-particle spectrum and two-body interactions.

\subsection{Two-band model in momentum space}
\label{sec:tbH}

To express the Bose-Hubbard Hamiltonian in momentum space, we Fourier expand
the operators via
$
c_{S i}^\dagger = \frac{1}{\sqrt{N_c}} \sum_\mathbf{k} 
e^{-i \mathbf{k} \cdot \mathbf{r_{S i}}} c_{S \mathbf{k}}^\dagger,
$
where $N_c$ is the number of unit cells in the lattice, $\mathbf{k}$ is the 
crystal momentum in the first Brillouin zone, and $\mathbf{r_{S i}}$ is the 
position of the site $Si$. Then a compact way to express the single-particle 
(hopping) term in general is
\begin{align}
\label{eqn:H0}
\mathcal{H}_0 = \sum_\mathbf{k}
\begin{pmatrix}
c_{A \mathbf{k}}^\dagger & c_{B \mathbf{k}}^\dagger 
\end{pmatrix}
\begin{pmatrix}
d_\mathbf{k}^0 + d_\mathbf{k}^z & d_\mathbf{k}^x - i d_\mathbf{k}^y \\
d_\mathbf{k}^x + i d_\mathbf{k}^y & d_\mathbf{k}^0 - d_\mathbf{k}^z
\end{pmatrix}
\begin{pmatrix}
c_{A \mathbf{k}} \\ c_{B \mathbf{k}}
\end{pmatrix},
\end{align}
where the field vector
$
\mathbf{d}_\mathbf{k} = \big(d^x_\mathbf{k}, d^y_\mathbf{k}, d^z_\mathbf{k}\big)
$
that is coupled to a vector of Pauli matrices
$
\boldsymbol{\sigma} = (\sigma_x, \sigma_y, \sigma_z)
$
governs the sublattice degrees of freedom.
Here $d_\mathbf{k}^0$, $d^x_\mathbf{k}$, $d^y_\mathbf{k}$ and $d^z_\mathbf{k}$ 
all depend on the specific details of the hopping parameters for a given lattice,
and we do not make any assumption on their $\mathbf{k}$ dependence.
Thus the Bloch bands are determined by the eigenvalues of the Hamiltonian 
matrix shown in Eq.~(\ref{eqn:H0}), leading to
$
\varepsilon_{s \mathbf{k}} = d_\mathbf{k}^0 + s d_\mathbf{k},
$
where $s = \pm$ labels the upper and lower bands, and 
$d_\mathbf{k} = |\mathbf{d}_\mathbf{k}|$ is the magnitude of the sublattice field.
The corresponding Bloch states $|s \mathbf{k} \rangle$ can be represented as
$
|+,\mathbf{k}\rangle = \left( u_\mathbf{k}, 
\; v_\mathbf{k} e^{i\varphi_\mathbf{k}} \right)^\mathrm{T}
$
and
$
|-,\mathbf{k}\rangle = \left( -v_\mathbf{k} e^{-i\varphi_\mathbf{k}}, 
\; u_\mathbf{k} \right)^\mathrm{T},
$
where
\begin{align}
&u_\mathbf{k} / v_\mathbf{k} = \sqrt{\frac{1}{2} \pm \frac{d^z_\mathbf{k}}{2d_\mathbf{k}}},
\\
&\varphi_\mathbf{k} = \arg(d^x_\mathbf{k} + i d^y_\mathbf{k}),
\end{align}
and $\mathrm{T}$ denotes a transpose.  

Similarly a compact way to express the interaction terms in general is
\begin{align}
\label{eqn:HI}
\mathcal{H}_I = \frac{1}{2N_c} \sum_{\substack{SS' \mathbf{k} \mathbf{k'} \mathbf{q}}}
U_{SS'}(\mathbf{q}) c_{S, \mathbf{k+q}}^\dagger c_{S', \mathbf{k'-q}}^\dagger
c_{S' \mathbf{k'}} c_{S \mathbf{k}},
\end{align}
where $\mathbf{q}$ is the exchanged momentum between particles, and
$
U_{SS'}(\mathbf{q}) = U_{SS'}^*(-\mathbf{q}) = U_{S'S}(-\mathbf{q})
$
by definition. Note that $U_{SS}(\mathbf{q})$ depends not only on the 
onsite interaction $U_S$ but also on $V_{Si, Si'}$.
Thus while the intra-sublattice interactions $U_{AA}(\mathbf{q})$ and 
$U_{BB}(\mathbf{q})$ are real and even functions of $\mathbf{q}$ in general, 
the inter-sublattice interaction $U_{AB}(\mathbf{q})$ can be complex.
For instance, in the case of an extended Bose-Hubbard model with 
only nearest-neighbor hopping $t$, onsite repulsion $U$ and nearest-neighbor 
repulsion $V$, they can be written as
$
U_{AA}(\mathbf{q}) = U_{BB}(\mathbf{q}) = U
$
and
$
U_{AB}(\mathbf{q}) =  V \sum_{\boldsymbol{\delta_{nn}}} 
e^{i\mathbf{q} \cdot \boldsymbol{\delta_{nn}}}
 = -(V/t) (d^x_\mathbf{q} - i d^y_\mathbf{q}),
$
where $\boldsymbol{\delta_{nn}}$ denotes the nearest neighbors of $A$ sublattice.
Therefore, in this particular case, $d^x_\mathbf{q} = d^x_{-\mathbf{q}}$ is necessarily 
an even function of $\mathbf{q}$ while $d^y_\mathbf{q} = -d^y_{-\mathbf{q}}$ 
is an odd one. For example 
$
\boldsymbol{\delta_{nn}} \in \{ (\pm a, 0), (0, \pm a) \}
$
and 
$U_{AB}(\mathbf{q}) = 2V[\cos(q_x a) + \cos(q_y a)]$ 
in a square lattice, but
$
\boldsymbol{\delta_{nn}} \in \{ (a, 0), (-a/2, \pm a\sqrt{3}/2) \}
$
and
$
U_{AB}(\mathbf{q}) = V[ \cos(q_x a) + 2\cos(q_x a/2) \cos(\sqrt{3}q_y a/2)]
- i V [\sin(q_x a) - 2\sin(q_x a/2) \cos(\sqrt{3}q_y a/2)] 
$
in a honeycomb lattice. 

The Bogoliubov spectrum for the total Hamiltonian 
$
\mathcal{H} = \mathcal{H}_0 + \mathcal{H}_I
$ 
can be obtained numerically through a straightforward application of the Bogoliubov 
approximation. Since the resultant Bogoliubov Hamiltonian matrix is $4 \times 4$, 
the spectrum has four branches, i.e., there are two quasiparticle and two 
quasihole bands that are related to each other through quasiparticle-quasihole 
symmetry. However, our main task here is to reveal a direct connection between 
the low-energy Bogoliubov modes and the quantum geometry of the Bloch states. 
Such a task can be achieved analytically by first projecting $\mathcal{H}$ 
onto the lower Bloch band, and then applying the Bogoliubov approximation to the 
projected Hamiltonian.

\subsection{Projection onto the lower Bloch band}
\label{sec:pH}

Suppose BEC takes place at the Bloch state $|-, \mathbf{k_c} \rangle$. 
Next we assume that the energy gap $2d_\mathbf{k_c}$ between the lower and upper 
Bloch bands near this ground state is much larger than the interaction 
energy, and project $\mathcal{H}$ to the lower band. 
Thus we first express $\mathcal{H}$ in the Bloch band basis, i.e.,
$
c_{S \mathbf{k}} = \sum_s \langle S | s \mathbf{k} \rangle c_{s \mathbf{k}},
$
and then discard those terms that involve the upper band, i.e.,
we set
$
c_{S \mathbf{k}} \to \langle S | -, \mathbf{k} \rangle c_{-, \mathbf{k}}
$
or more explicitly
$
c_{A \mathbf{k}} \to -v_\mathbf{k} e^{-i\varphi_\mathbf{k}} c_{-, \mathbf{k}}
$
and
$
c_{B \mathbf{k}} \to u_\mathbf{k} c_{-, \mathbf{k}}.
$
Here $c_{s \mathbf{k}}$ annihilates a particle from the Bloch state 
$|s \mathbf{k} \rangle$. This procedure leads to the projected Hamiltonian
\begin{align}
\label{eqn:HP}
\mathcal{H}_P &= \sum_\mathbf{k} \big( \varepsilon_{-,\mathbf{k}} - \mu \big) c_{-, \mathbf{k}}^\dagger  c_{-, \mathbf{k}}
\nonumber \\
&+ \frac{1}{2N_c} \sum_{\mathbf{k}\mathbf{k'}\mathbf{q}}
f_{\mathbf{k+q}, \mathbf{k'-q}}^{\mathbf{k'}, \mathbf{k}} 
c_{-, \mathbf{k+q}}^\dagger c_{-, \mathbf{k'-q}}^\dagger
c_{-, \mathbf{k'}} c_{-, \mathbf{k}},
\end{align}
where the second term describes the dressed density-density interactions in 
the lower Bloch band with the effective interaction amplitude
$
f_{\mathbf{k+q}, \mathbf{k'-q}}^{\mathbf{k'}, \mathbf{k}} 
= \sum_{SS'} U_{SS'}(\mathbf{q}) 
\langle -, \mathbf{k+q}|S \rangle
\langle -, \mathbf{k'-q}|S' \rangle
\langle S' | -, \mathbf{k'} \rangle
\langle S | -, \mathbf{k} \rangle.
$
In terms of the Bloch factors, this effective interaction becomes
\begin{align}
f_{\mathbf{k+q}, \mathbf{k'-q}}^{\mathbf{k'}, \mathbf{k}} 
&= U_{AA}(\mathbf{q}) v_\mathbf{k+q} v_\mathbf{k'-q} v_\mathbf{k'} v_\mathbf{k}
e^{i(\varphi_\mathbf{k+q}+\varphi_\mathbf{k'-q}-\varphi_\mathbf{k'}-\varphi_\mathbf{k})}
\nonumber\\
&+ U_{BB}(\mathbf{q}) u_\mathbf{k+q} u_\mathbf{k'-q} u_\mathbf{k'} u_\mathbf{k}
\nonumber\\
&+U_{AB}(\mathbf{q}) v_\mathbf{k+q} u_\mathbf{k'-q} u_\mathbf{k'} v_\mathbf{k}
e^{i(\varphi_\mathbf{k'+q}-\varphi_\mathbf{k})}
\nonumber\\
&+U_{BA}(\mathbf{q}) u_\mathbf{k+q} v_\mathbf{k'-q} v_\mathbf{k'} u_\mathbf{k}
e^{i(\varphi_\mathbf{k'-q}-\varphi_\mathbf{k'})}.
\label{eqn:feff}
\end{align}
Note that the $\mathbf{q}$ dependence of 
$
f_{\mathbf{k+q}, \mathbf{k'-q}}^{\mathbf{k'}, \mathbf{k}} 
$
remains there even in the absence of the offsite interaction term $V_{S i, S'i'}$ 
in Eq.~(\ref{eqn:BHHam}).

Equation~(\ref{eqn:HP}) is expected to be quantitatively accurate in describing
the low-energy physics when the occupation of the upper band is negligible, 
e.g., in the weakly-interacting limit. 
For instance, in the case of spin-orbit-coupled Bose gases, the corresponding 
$\mathcal{H}_P$ works surprisingly well as it perfectly reproduces the Bogoliubov 
spectrum not only at low momenta but also near the band touchings~\cite{subasi22}.
See also the App. A below.

\section{Bogoliubov Theory}
\label{sec:Bt}

In the Bogoliubov approximation, we first replace the creation and annihilation 
operators in accordance with
$
c_{-, \mathbf{k}} = \sqrt{N_0} \delta_{\mathbf{k} \mathbf{k_c}} + \tilde{c}_{-, \mathbf{k}},
$
where $N_0$ is the number of condensed particles at the Bloch state 
$|-, \mathbf{k_c} \rangle$, $\delta_{\mathbf{k} \mathbf{k_c}}$ is a Kronecker delta,
and the operator $\tilde{c}_{-, \mathbf{k}}$ denotes the fluctuations on top of 
the many-body ground state. Then we set the first-order fluctuations to zero, 
and discard the third- and fourth-order fluctuations. 
The former condition gives
$\mu = \varepsilon_{-,\mathbf{k_c}} + n_0 f_{\mathbf{k_c} \mathbf{k_c}}^{\mathbf{k_c} \mathbf{k_c}},
$
leading to
\begin{align}
\mu = \varepsilon_{-,\mathbf{k_c}} + n_0\big[ 
&U_{AA}(\mathbf{0}) v_\mathbf{k_c}^4 + U_{BB}(\mathbf{0}) u_\mathbf{k_c}^4 
\nonumber \\
&+ 2U_{AB}(\mathbf{0}) u_\mathbf{k_c}^2 v_\mathbf{k_c}^2 
\big],
\label{eqn:mu}
\end{align}
where 
$
U_{AB}(\mathbf{0}) = U_{BA}(\mathbf{0})  
$
is real by definition and $n_0 = N_0/N_c$ is the condensate filling per unit cell. 
Note that the condensate filling within a unit cell (i.e., on sublattices 
A and B) is not necessarily uniform unless 
$u_\mathbf{k_c} = v_\mathbf{k_c} = 1/\sqrt{2}$ 
(i.e., $d_\mathbf{k_c}^z = 0$) is favored by the interactions.

\subsection{Bogoliubov Hamiltonian}
\label{sec:BH}

The second-order fluctuations are described by the Bogoliubov Hamiltonian
\begin{align} 
\label{eqn:BH}
\mathcal{H}_\mathrm{B} &= \frac{1}{2} \sum_\mathbf{q}^\prime
\begin{pmatrix}
\tilde{c}_{-, \mathbf{k_c+q}}^\dagger & \tilde{c}_{-, \mathbf{k_c-q}} 
\end{pmatrix}
\begin{pmatrix}
h^{pp}_\mathbf{q} & h^{ph}_\mathbf{q} \\
h^{hp}_\mathbf{q} & h^{hh}_\mathbf{q} 
\end{pmatrix}
\begin{pmatrix}
\tilde{c}_{-, \mathbf{k_c+q}} \\ \tilde{c}_{-, \mathbf{k_c-q}}^\dagger 
\end{pmatrix},
\end{align}
where the diagonal elements 
$
h^{hh}_\mathbf{q} = h^{pp}_{-\mathbf{q}}
$
are given by
$
h^{pp}_\mathbf{q} = \varepsilon_{-,\mathbf{k_c+q}} - \mu + \frac{n_0}{2} 
\big(f_{\mathbf{k_c}, \mathbf{k_c+q}}^{\mathbf{k_c}, \mathbf{k_c+q}} 
+ f_{\mathbf{k_c+q}, \mathbf{k_c}}^{\mathbf{k_c+q}, \mathbf{k_c}}
+ f_{\mathbf{k_c+q}, \mathbf{k_c}}^{\mathbf{k_c}, \mathbf{k_c+q}} 
+ f_{\mathbf{k_c}, \mathbf{k_c+q}}^{\mathbf{k_c+q}, \mathbf{k_c}} 
\big),
$
and the off-diagonal elements
$
h^{hp}_\mathbf{q}  = (h^{ph}_\mathbf{q})^*.
$
are given by
$
h^{ph}_\mathbf{q} = \frac{n_0}{2}
\big(f_{\mathbf{k_c+q}, \mathbf{k_c-q}}^{\mathbf{k_c}, \mathbf{k_c}} 
+ f_{\mathbf{k_c-q}, \mathbf{k_c+q}}^{\mathbf{k_c}, \mathbf{k_c}} 
\big).
$
The prime symbol in Eq.~(\ref{eqn:BH}) indicates that the summation excludes 
the condensed state. In terms of the Bloch factors, the matrix elements become
\begin{align}
h^{pp}_\mathbf{q} &= \varepsilon_{-,\mathbf{k_c+q}} - \mu + n_0 \big\lbrace
\big[U_{AA}(\mathbf{q}) +U_{AA}(\mathbf{0})\big] v_\mathbf{k_c}^2 v_\mathbf{k_c+q}^2  
\nonumber \\
& + \big[U_{BB}(\mathbf{q}) +U_{BB}(\mathbf{0})\big] u_\mathbf{k_c}^2 u_\mathbf{k_c+q}^2
\nonumber \\
& + U_{AB}(\mathbf{0}) \big(u_\mathbf{k_c}^2 v_\mathbf{k_c+q}^2 + v_\mathbf{k_c}^2 u_\mathbf{k_c+q}^2\big) 
\nonumber \\
& + 2 \mathrm{Re}\big[U_{AB}(\mathbf{q}) e^{i(\varphi_\mathbf{k_c+q} - \varphi_\mathbf{k_c})}\big]
u_\mathbf{k_c} v_\mathbf{k_c} u_\mathbf{k_c+q} v_\mathbf{k_c+q}
\big\rbrace,
\label{eqn:hpp}
\\
h^{ph}_\mathbf{q} &= n_0\big[
U_{AA}(\mathbf{q}) 
e^{i(\varphi_\mathbf{k_c+q} + \varphi_\mathbf{k_c-q} - 2\varphi_\mathbf{k_c})}
v_\mathbf{k_c}^2 v_\mathbf{k_c+q} v_\mathbf{k_c-q} 
\nonumber \\
& + U_{BB}(\mathbf{q}) u_\mathbf{k_c}^2 u_\mathbf{k_c+q} u_\mathbf{k_c-q}
\nonumber \\
& + U_{AB}(\mathbf{q}) e^{i(\varphi_\mathbf{k_c+q} - \varphi_\mathbf{k_c})}
u_\mathbf{k_c} v_\mathbf{k_c} u_\mathbf{k_c-q} v_\mathbf{k_c+q}
\nonumber \\
& + U_{AB}(-\mathbf{q}) e^{i(\varphi_\mathbf{k_c-q} - \varphi_\mathbf{k_c})}
u_\mathbf{k_c} v_\mathbf{k_c} u_\mathbf{k_c+q} v_\mathbf{k_c-q}
\big],
\label{eqn:hph}
\end{align}
for the particle-particle and particle-hole sectors, where $\mathrm{Re}$ 
denotes the real part.

The Bogoliubov spectrum $E_{s\mathbf{q}}$ for Eq.~(\ref{eqn:BH}) is determined 
by the eigenvalues of $\tau_z \mathbf{h}_\mathbf{q}$ so that the bosonic 
commutation rules are properly taken into account, where $\tau_z$ is the 
third Pauli matrix in the particle-hole space and $\mathbf{h}_\mathbf{q}$ 
is the Hamiltonian matrix shown in Eq.~(\ref{eqn:BH}). 
Thus the spectrum has two modes for a given $\mathbf{q}$, i.e.,
\begin{align}
\label{eqn:Esq}
&E_{s \mathbf{q}} = A_\mathbf{q} + s \sqrt{B_\mathbf{q}^2 - |h^{ph}_\mathbf{q}|^2},
\\
&A_\mathbf{q} / B_\mathbf{q} = \frac{h^{pp}_\mathbf{q} \mp h^{hh}_\mathbf{q}}{2},
\end{align}
where $s = \pm$ denotes, respectively, the quasiparticle and quasihole branches
in the first line. 
Here $A_\mathbf{q}$ is odd in $\mathbf{q}$, and $B_\mathbf{q}$ and 
$h^{ph}_\mathbf{q}$ are even in $\mathbf{q}$, so that the quasiparticle-quasihole 
symmetry 
$
E_{+, \mathbf{q}} = -E_{-, -\mathbf{q}}
$
manifests in the spectrum.

\subsection{Low-energy Bogoliubov excitations}
\label{sec:Bs}

Since our primary objective is to derive an analytical expression for the 
low-energy Bogoliubov modes, next we calculate $E_{s \mathbf{q}}$ accurately 
up to first order in $\mathbf{q}$. For this purpose we first recall that
$
U_{AB}(\mathbf{q}) = U_{AB}^*(-\mathbf{q}),
$ 
and therefore $U_{AB}(\mathbf{0})$ is always real. Furthermore, given that 
the zeroth-order contribution from the imaginary part
$
\mathrm{Im}[h^{ph}_\mathbf{0}] = 0
$
vanishes, its second-order contribution (which contributes to the square-root
term in $E_{s \mathbf{q}}$ at quartic order in $\mathbf{q}$) is not needed for the 
determination of the effective-mass tensor of the superfluid carriers. 
Thus we may simply substitute
$
|h^{ph}_\mathbf{q}|^2 \to C_\mathbf{q}^2
$
for the low-$\mathbf{q}$ modes, where
\begin{align}
C_\mathbf{q} &= n_0\big\lbrace
v_\mathbf{k_c}^2 v_\mathbf{k_c+q} v_\mathbf{k_c-q} U_{AA}(\mathbf{q}) 
+ u_\mathbf{k_c}^2 u_\mathbf{k_c+q} u_{k_c-q} U_{BB}(\mathbf{q}) 
\nonumber \\ 
&+ u_\mathbf{k_c} v_\mathbf{k_c} u_\mathbf{k_c-q} v_\mathbf{k_c+q} 
\mathrm{Re} \big[U_{AB}(\mathbf{q}) e^{i(\varphi_\mathbf{k_c+q} - \varphi_\mathbf{k_c})}\big]
\nonumber \\ 
&+ u_\mathbf{k_c} v_\mathbf{k_c} u_\mathbf{k_c+q} v_\mathbf{k_c-q} 
\mathrm{Re} \big[U_{AB}(-\mathbf{q}) e^{i(\varphi_\mathbf{k_c-q} - \varphi_\mathbf{k_c})} \big]
\big\rbrace
\end{align}
is taken as real up to second-order in $\mathbf{q}$. Note here that, since
$
\varphi_\mathbf{k_c+q} + \varphi_\mathbf{k_c-q} - 2\varphi_\mathbf{k_c}
$
is even in $\mathbf{q}$ and vanishes at the zeroth order, the $\mathbf{q}$ 
dependence coming from its cosine is at least quartic order, and therefore 
it is dropped from the first term as well.
Then we only need the expansion of
$
A_\mathbf{q} = \sum_\ell A_\ell q_\ell + O(\mathbf{q}^3),
$
up to first order in $\mathbf{q}$, and the expansions of 
$
B_\mathbf{q} = B_\mathbf{0} + (1/2) \sum_{\ell \ell'} B_{\ell \ell'} q_\ell q_{\ell'} + O(\mathbf{q}^4)
$
and
$
C_\mathbf{q} = C_\mathbf{0} + (1/2) \sum_{\ell \ell'} C_{\ell \ell'} q_\ell q_{\ell'} + O(\mathbf{q}^4)
$
up to second orders in $\mathbf{q}$. Here $q_\ell$ refers to the $\ell$th component
of the $\mathbf{q} = (q_x, q_y, q_z)$ vector,
$
A_\ell = (\partial A_\mathbf{q} / \partial q_\ell)_\mathbf{q = 0},
$
$
B_{\ell \ell'}  = (\partial^2 B_\mathbf{q} / \partial q_\ell \partial q_{\ell'})_\mathbf{q = 0},
$
and similarly
$
C_{\ell \ell'}  = (\partial^2 C_\mathbf{q} / \partial q_\ell \partial q_{\ell'})_\mathbf{q = 0}.
$

The zeroth-order coefficients
$
B_\mathbf{0} = \varepsilon_\mathbf{k_c} - \mu + 2n_0\big[v_\mathbf{k_c}^4 U_{AA}(\mathbf{0}) 
+ u_\mathbf{k_c}^4 U_{BB}(\mathbf{0}) + u_\mathbf{k_c}^2 v_\mathbf{k_c}^2 U_{AB}(\mathbf{0})\big]
$
and
$
C_\mathbf{0} = n_0\big[v_\mathbf{k_c}^4 U_{AA}(\mathbf{0}) + u_\mathbf{k_c}^4 U_{BB}(\mathbf{0})
+ 2 u_\mathbf{k_c}^2 v_\mathbf{k_c}^2 U_{AB}(\mathbf{0})\big]
$
are equal to each other due to Eq.~(\ref{eqn:mu}), which guarantees that the 
$\mathbf{q = 0}$ mode is gapless, i.e., $E_{s\mathbf{0}} = 0$. Thus we find
\begin{align}
\label{eqn:Elow}
E_{s \mathbf{q}} = \sum_\ell A_\ell q_\ell + s \sqrt{
B_\mathbf{0} \sum_{\ell \ell'} \big(B_{\ell \ell'} - C_{\ell \ell'}\big) q_\ell q_{\ell'}
} 
+ O(\mathbf{q}^2)
\end{align}
for the low-energy quasiparticle/quasihole excitations, where 
\begin{align}
A_\ell & = \dot{\varepsilon}_\mathbf{k_c}^\ell + n_0\big[
4U_{AA}(\mathbf{0}) v_\mathbf{k_c}^3 \dot{v}_\mathbf{k_c}^\ell 
+4U_{BB}(\mathbf{0}) u_\mathbf{k_c}^3 \dot{u}_\mathbf{k_c}^\ell 
\nonumber\\
&+ \dot{U}_{AA}^\ell(\mathbf{0}) v_\mathbf{k_c}^4 + \dot{U}_{BB}^\ell(\mathbf{0}) u_\mathbf{k_c}^4 
+ 2 \dot{U}_{AB}^{\ell} (\mathbf{0}) u_\mathbf{k_c}^2 v_\mathbf{k_c}^2
\nonumber \\
&+ 4U_{AB}(\mathbf{0}) u_\mathbf{k_c} v_\mathbf{k_c}
	\big(u_\mathbf{k_c} \dot{v}_\mathbf{k_c}^\ell + v_\mathbf{k_c} \dot{u}_\mathbf{k_c}^\ell \big)
\big]
\label{eqn:All}
\end{align}
is the coefficient of the linear term, and 
\begin{widetext}
\begin{align}
B_{\ell \ell'}-C_{\ell \ell'}  = (M^{-1})_{\ell\ell'} + n_0\big\lbrace
& 2 U_{AA}(\mathbf{0}) v_\mathbf{k_c}^2 
\big(v_\mathbf{k_c} \ddot{v}_{\mathbf{k_c}}^{\ell\ell'} + 3\dot{v}_\mathbf{k_c}^\ell  \dot{v}_\mathbf{k_c}^{\ell'}\big)
+ 2 U_{BB}(\mathbf{0}) u_\mathbf{k_c}^2  
\big(u_\mathbf{k_c} \ddot{u}_{\mathbf{k_c}}^{\ell\ell'} + 3\dot{u}_\mathbf{k_c}^\ell  \dot{u}_\mathbf{k_c}^{\ell'}\big)
\nonumber \\
+ &2\big[\dot{U}_{AA}^\ell(\mathbf{0}) \dot{v}_\mathbf{k_c}^{\ell'}
  +\dot{U}_{AA}^{\ell'}(\mathbf{0}) \dot{v}_\mathbf{k_c}^{\ell} \big] v_\mathbf{k_c}^3 
+ 2 \big[ \dot{U}_{BB}^\ell(\mathbf{0}) \dot{u}_\mathbf{k_c}^{\ell'}
	+\dot{U}_{BB}^{\ell'}(\mathbf{0}) \dot{u}_\mathbf{k_c}^{\ell} \big] u_\mathbf{k_c}^3 
\nonumber\\
+ &2U_{AB}(\mathbf{0}) \big[
 u_\mathbf{k_c}^2\big( v_\mathbf{k_c} \ddot{v}_\mathbf{k_c}^{\ell \ell'}
 	+ \dot{v}_\mathbf{k_c}^\ell \dot{v}_\mathbf{k_c}^{\ell'}\big)
+  v_\mathbf{k_c}^2\big( u_\mathbf{k_c} \ddot{u}_\mathbf{k_c}^{\ell \ell'} 
 	+ \dot{u}_\mathbf{k_c}^\ell \dot{u}_\mathbf{k_c}^{\ell'}\big) 
+ 2u_\mathbf{k_c} v_\mathbf{k_c}\big(\dot{u}_\mathbf{k_c}^\ell \dot{v}_\mathbf{k_c}^{\ell'}
	+ \dot{v}_\mathbf{k_c}^\ell \dot{u}_\mathbf{k_c}^{\ell'}\big)
	\big]
\nonumber \\	
+ &4\mathrm{Re}\big[\dot{U}_{AB}^\ell (\mathbf{0}) \dot{u}_\mathbf{k_c}^{\ell'} 
	+ \dot{U}_{AB}^{\ell'} (\mathbf{0}) \dot{u}_\mathbf{k_c}^{\ell}\big] u_\mathbf{k_c}v_\mathbf{k_c}^2 
\big\rbrace
\label{eqn:Bll}
\end{align}
\end{widetext}
is the coefficient of the quadratic term inside the square root. Here
$
(M^{-1})_{\ell\ell'} = (\partial^2 \varepsilon_{-,\mathbf{k_c+q}} / \partial q_\ell \partial q_{\ell'})_{\mathbf{q = 0}}
$
is the matrix element of the inverse band-mass tensor $\mathbf{M^{-1}}$ 
for a particle in the lower Bloch band, 
$
\dot{u}_\mathbf{k_c}^\ell = (\partial u_\mathbf{k_c+q}/\partial q_\ell)_\mathbf{q = 0},
$
$
\ddot{u}_\mathbf{k_c}^{\ell \ell'} = (\partial^2 u_\mathbf{k_c+q}
/\partial q_\ell \partial q_{\ell'})_\mathbf{q = 0}
$
and
$
\dot{U}_{SS'}^\ell(\mathbf{0}) =  
[\partial U_{SS'}(\mathbf{q}) / \partial q_\ell]_{\mathbf{q = 0}}.
$
Note that Eq.~(\ref{eqn:Bll}) can be interpreted as the inverse effective-mass 
tensor for the superfluid carriers dressed by the presence of an upper 
Bloch band~\cite{subasi22}.

Equations~(\ref{eqn:All}) and~(\ref{eqn:Bll}) can be simplified considerably 
as follows. Since $U_{AB}(\mathbf{q}) = U_{AB}^*(-\mathbf{q})$, we first 
note that
$
\mathrm{Re}[U_{AB}(\mathbf{q})] = \mathrm{Re}[U_{AB}^*(-\mathbf{q})]
$
is an even function of $\mathbf{q}$, and therefore take
$
\mathrm{Re}[\dot{U}_{AB}^\ell(\mathbf{0})] = 0.
$
Similarly $U_{SS}(\mathbf{q})$ is also an even function of $\mathbf{q}$, 
and therefore take $\dot{U}_{SS}^\ell(\mathbf{0}) = 0$.
Furthermore we suppose 
$
U_{AA}(\mathbf{0}) = U_{BB}(\mathbf{0}) = \mathcal{U}
$
are equal for both sublattices,
$
U_{AB}(\mathbf{0}) = \mathcal{V},
$
and
$
\dot{\varepsilon}_\mathbf{k_c}^\ell = 0.
$
It is also convenient to substitute
$
u_\mathbf{q} = \cos(\theta_\mathbf{q}/2)
$
and
$
v_\mathbf{q} = \sin(\theta_\mathbf{q}/2)
$
without loss of generality as they satisfy
$
u_\mathbf{q}^2 + v_\mathbf{q}^2 = 1.
$
Note that $\theta_\mathbf{q}$ and $\varphi_\mathbf{q}$ correspond, respectively, 
to the azimuthal and polar angles on the Bloch sphere. 
With these simplifications, we find that
\begin{align}
\label{eqn:Al}
A_\ell &= - \frac{n_0}{2} \mathcal{U} \sin(2\theta_\mathbf{k_c}) \dot{\theta}_\mathbf{k_c}^\ell,
\\
\label{eqn:B0}
B_\mathbf{0} &= n_0 \mathcal{U} + \frac{n_0}{2} \big(\mathcal{V}-\mathcal{U}\big) \sin^2\theta_\mathbf{k_c},
\\
\label{eqn:BlCl}
B_{\ell \ell'}-C_{\ell \ell'}  &= (M^{-1})_{\ell\ell'} + \frac{n_0}{4}
\big(\mathcal{V} - \mathcal{U}\big) 
\nonumber \\
&\times \big[
\sin(2\theta_\mathbf{k_c}) \ddot{\theta}_\mathbf{k_c}^{\ell \ell'} 
+ 2\cos(2\theta_\mathbf{k_c}) 
	\dot{\theta}_\mathbf{k_c}^\ell \dot{\theta}_\mathbf{k_c}^{\ell'}
\big]
\end{align}
are the desired expansion coefficients in general, where
$
\dot{\theta}_\mathbf{k_c}^\ell = (\partial \theta_\mathbf{k_c+q}/\partial q_\ell)_\mathbf{q = 0}
$
and
$
\ddot{\theta}_\mathbf{k_c}^{\ell \ell'} = (\partial^2 \theta_\mathbf{k_c+q}
/\partial q_\ell \partial q_{\ell'})_\mathbf{q = 0}.
$
Equation~(\ref{eqn:BlCl}) reveals that the dressing of the effective-mass tensor 
is caused by the presence of a second band in the Bloch spectrum, and that it 
has a peculiar dependence on the geometry of the Bloch sphere.
Note that Eqs.~(\ref{eqn:Al}),~(\ref{eqn:B0}) and~(\ref{eqn:BlCl}) do not depend 
on $\varphi_\mathbf{q}$. See also a related discussion at the end of Sec. III in 
Ref.~\cite{julku22}. 

It is important to emphasize that these generic expressions are valid and 
applicable to a broad range of two-band Bose-Hubbard models.
In the particular case when $\theta_\mathbf{k_c} = \pi/2$, i.e., 
when $d_\mathbf{k_c}^z = 0$, Eq.~(\ref{eqn:Elow}) can be written as
\begin{align}
\label{eqn:Epio2}
&E_{s \mathbf{q}} = s \sqrt{
n_0 \frac{\mathcal{U}+\mathcal{V}}{2} \sum_{\ell \ell'}
(M_\textrm{eff}^{-1})_{\ell\ell'} 
q_\ell q_{\ell'}
}
+ O(\mathbf{q}^2), 
\\
\label{eqn:Meff}
&(M_\textrm{eff}^{-1})_{\ell\ell'} = (M^{-1})_{\ell\ell'} 
+ n_0 \frac{\mathcal{U} - \mathcal{V}}{2} 
\dot{\theta}_\mathbf{k_c}^\ell \dot{\theta}_\mathbf{k_c}^{\ell'},
\end{align}
where 
$
(M_\textrm{eff}^{-1})_{\ell\ell'} 
$
is the matrix element of the inverse effective-mass tensor $\mathbf{M_{eff}^{-1}}$
for the superfluid carriers.
This case corresponds to a uniform condensate filling on sublattices A 
and B since $u_\mathbf{k_c} = v_\mathbf{k_c} = 1/\sqrt{2}$. Given that both 
sublattices are equally populated, $n_0/2$ corresponds to the condensate filling 
per lattice site in the system, and hence to the proper prefactor 
for the effective mass.
For instance, when $\mathcal{U} > \mathcal{V}$, the ground state of a 
flat-band BEC is expected to be uniform over the unit cell as this 
configuration minimizes the repulsive interactions~\cite{you12}.
We note in passing that since $\dot{\theta}_\mathbf{q}^{\ell}$ is trivially zero 
when $d_\mathbf{q}^z = 0$ for every $\mathbf{q}$ in the entire Brilluoin zone, 
the presence of a finite geometric contribution relies on a non-trivial 
$d_\mathbf{q}^z$ to begin with. For instance, in the case of bipartite lattices, 
next-nearest-neighbor hopping processes may give rise to such an intra-sublattice 
term in the Bloch Hamiltonian. See Appendix B for example models. 
Unless $d_\mathbf{q}^z$ is coupled with 
a $d_\mathbf{q}^x$ and/or $d_\mathbf{q}^y$ term in the Bloch Hamiltonian, 
$\theta_\mathbf{q} = \{0, \pi\}$ for every $\mathbf{q}$ in the entire 
Brilluoin zone, and therefore, $\dot{\theta}_\mathbf{k_c}^{\ell} = 0$ becomes trivial.
Furthermore the geometric dressing is also trivial when $\mathcal{U} = \mathcal{V}$, 
whose physical significance is not obvious.

On the other hand when $\theta_\mathbf{k_c} = \{0, \pi\}$, i.e., 
when $d_\mathbf{k_c}^x = 0 = d_\mathbf{k_c}^y$ and $d_\mathbf{k_c}^z \gtrless 0$, Eq.~(\ref{eqn:Elow}) reduces to
$
E_{s \mathbf{q}} = s \sqrt{n_0 \mathcal{U} \sum_{\ell \ell'} \big[
(M^{-1})_{\ell\ell'} + n_0 \frac{\mathcal{V}-\mathcal{U}}{2} 
\dot{\theta}_\mathbf{k_c}^\ell \dot{\theta}_\mathbf{k_c}^{\ell'}
\big] q_\ell q_{\ell'}
}
+ O(\mathbf{q}^2)
$.
While the $\theta_\mathbf{k_c} = 0$ case with $u_\mathbf{k_c} = 1$ and $v_\mathbf{k_c} = 0$ 
corresponds to a condensate filling that is entirely 
on sublattice B, $\theta_\mathbf{k_c} = \pi$ case with $u_\mathbf{k_c} = 0$ 
and $v_\mathbf{k_c} = 1$ corresponds to a condensate filling that is entirely 
on sublattice A. Given that one of the sublattices is empty, $n_0$ corresponds 
to the condensate filling per site for the occupied sublattice, which explains 
the difference between the prefactor of the effective mass here and 
in Eq.~(\ref{eqn:Epio2}).
Thus, since the condensate filling has the structure of a charge-density-wave pattern 
in both of these extreme cases, the inter-sublattice interaction $\mathcal{V}$ must 
disappear from the Bogoliubov spectrum because one of the sublattices is not 
macroscopically occupied. In fact it can be shown that 
$
\dot{\theta}_\mathbf{k_c}^\ell = 2 \dot{v}_\mathbf{k_c}^\ell/ u_\mathbf{k_c} = 0
$
in general for both of these extreme cases, which makes their geometric dressing 
trivial. For instance, in the case of an extended Bose-Hubbard model that is discussed 
in App. A, such a density-wave superfluid (i.e., a supersolid) may occur 
when the nearest-neighbor repulsion $\mathcal{V}$ is sufficiently larger than the 
onsite one on a bipartite lattice, and the asymmetric occupation of the sublattices 
can become as dramatic only in the $\mathcal{V} \gg \mathcal{U}$ 
limit~\cite{danshita09, iskin11}.

We emphasize that the bare band-mass tensor in Eq.~(\ref{eqn:Meff}) and its 
dressing terms have completely different physical origins. While the usual term 
$(M^{-1})_{\ell\ell'}$ is associated with the intraband processes within 
the lower Bloch band in which the BEC occurs, the dressing terms 
are related to the interband processes that are induced by the interactions. 
That is why they have an overall factor of 
$\mathcal{U}$ and/or $\mathcal{V}$ in the front. 
The interband terms are quite peculiar because they depend not only on the 
Bloch bands but also on the Bloch states themselves, i.e., on the geometry
of the Bloch sphere. 
For this reason they are claimed to have a quantum-geometric 
origin~\cite{julku21a, julku21b, julku22, subasi22}. 
Their critical roles in Eqs.~(\ref{eqn:BlCl}) and~(\ref{eqn:Epio2}) are
to renormalize and dress the inverse effective-mass tensor of the superfluid 
carriers. Next we show that the geometric contribution of Eq.~(\ref{eqn:Meff}) 
can be related to the quantum-metric tensor of the underlying Bloch states 
under some specific conditions.

\subsection{Connection to the quantum-metric tensor}
\label{sec:qm}

The quantum-metric tensor corresponds to the real part of the quantum-geometric 
tensor~\cite{resta11}. For instance, in the case of a multiband Bloch Hamiltonian, 
it can be expressed in general as
$
g^{s\mathbf{k}}_{\ell\ell'} = \mathrm{Re} 
\big[
(\partial \langle s\mathbf{k} | / \partial k_\ell)
(\mathbb{I} - |s\mathbf{k} \rangle \langle s\mathbf{k}|)
(\partial |s\mathbf{k} \rangle / \partial k_{\ell'})
\big],
$
where $|s \mathbf{k} \rangle$ corresponds to the Bloch state for band $s$ at 
momentum $\mathbf{k}$, and
$
\mathbb{I} = \sum_s | s\mathbf{k} \rangle  \langle s\mathbf{k} |
$ 
denotes the identity operator for a given $\mathbf{k}$. 
In the case of two-band lattices where $s = \pm$, it can be shown that
$
g^{+, \mathbf{k}}_{\ell\ell'} = g^{-, \mathbf{k}}_{\ell\ell'} = 
g^{\mathbf{k}}_{\ell\ell'}
$
is given by
\begin{align}
\label{eqn:qm}
g^{\mathbf{k}}_{\ell\ell'} = 
\frac{1}{4} \dot{\theta}_\mathbf{k}^{\ell} \dot{\theta}_\mathbf{k}^{\ell'} + 
\frac{\sin^2\theta_\mathbf{k}}{4} \dot{\varphi}_\mathbf{k}^{\ell} \dot{\varphi}_\mathbf{k}^{\ell'},
\end{align}
where 
$
\dot{\varphi}_\mathbf{k}^{\ell} = \partial \varphi_\mathbf{k}/\partial k_\ell.
$
Thus the so-called quantum distance
$
ds^2 = \sum_{\ell \ell'} g^{\mathbf{k}}_{\ell\ell'} d k_\ell d k_{\ell'} = 
(d\theta_\mathbf{k} d\theta_\mathbf{k} 
+ \sin^2\theta_\mathbf{k} d\varphi_\mathbf{k} d\varphi_\mathbf{k})/4
$
clearly illustrates that $g^{\mathbf{k}}_{\ell\ell'}$ corresponds to nothing 
but to the natural Fubini-Study metric on the Bloch sphere with radius $r = 1/2$
~\footnote{
Note that the imaginary part of the quantum-geometric tensor
$
\Omega^{\pm, \mathbf{k}}_{\ell\ell'} = 
\pm \frac{\sin\theta_\mathbf{k}}{4} \big( \dot{\theta}_\mathbf{k}^{\ell} \dot{\varphi}_\mathbf{k}^{\ell'} 
+ \dot{\theta}_\mathbf{k}^{\ell'} \dot{\varphi}_\mathbf{k}^{\ell} \big)
$
is directly related to the Berry curvature
$
F^{\pm}_{\theta_\mathbf{k} \varphi_\mathbf{k}} = \mp \frac{1}{2}\sin\theta_\mathbf{k}
$
}.
When $\theta_\mathbf{k} \in \{0, \pi\}$ or $\dot{\varphi}_\mathbf{k}^{\ell} = 0$, Eq.~(\ref{eqn:qm}) reduces to
$
\dot{\theta}_\mathbf{k}^{\ell} \dot{\theta}_\mathbf{k}^{\ell'} / 4,
$
which gives precisely the interband contribution to the effective-mass tensor 
in Eq.~(\ref{eqn:Meff}) up to a prefactor, i.e., 
\begin{align}
\label{eqn:Meffg}
(M_\textrm{eff}^{-1})_{\ell\ell'} = (M^{-1})_{\ell\ell'} 
+ 2n_0 \big(\mathcal{U} - \mathcal{V}\big) 
g^{\mathbf{k_c}}_{\ell\ell'}.
\end{align}
However, since the former two cases have trivial geometry, here we concentrate 
only on the latter case (i.e., $\dot{\varphi}_\mathbf{k_c}^{\ell} = 0$) 
requiring that either (i)
$
d_\mathbf{k_c}^x = 0 = \dot{d}_\mathbf{k_c}^{x,\ell}
$ 
or (ii)
$
d_\mathbf{k_c}^y = 0 = \dot{d}_\mathbf{k_c}^{y,\ell},
$ 
but not both simultaneously, where
$
\dot{d}_\mathbf{k_c}^{i,\ell} = 
(\partial d_\mathbf{k}^i/\partial k_\ell)_{\mathbf{k} = \mathbf{k_c}}.
$
Note that Eq.~(\ref{eqn:Epio2}) is derived for a uniformly-condensed Bose gas, 
i.e.,
it assumes
$
d_\mathbf{k_c}^z = 0
$
as well.
Thus the latter possibility (ii) is in complete agreement with the previous 
works~\cite{julku21a, julku21b, julku22, subasi22}:
the quantum-metric tensor appears in the Bogoliubov spectrum of a uniformly-condensed 
Bose gas when all of the $\mathbf{k}$ states for the lowest-lying Bloch band 
$|-,\mathbf{k} \rangle$ admit real representation in the Brillouin zone, i.e., 
when $d_\mathbf{k}^y = 0$ for every $\mathbf{k}$. 
When this condition holds, it automatically guarantees that 
$\dot{d}_\mathbf{k}^{y,\ell} = 0$ for every $\mathbf{k}$ as well, 
leading eventually to
$
g^{\mathbf{k_c}}_{\ell\ell'} = \dot{d}_\mathbf{k_c}^{z,\ell} \dot{d}_\mathbf{k_c}^{z,\ell'} 
/ (2d_\mathbf{k_c}^{x})^2
$ 
for case (ii). Similarly we find
$
g^{\mathbf{k_c}}_{\ell\ell'} = \dot{d}_\mathbf{k_c}^{z,\ell} \dot{d}_\mathbf{k_c}^{z,\ell'} 
/ (2d_\mathbf{k_c}^{y})^2
$ 
for the former possibility (i) when $d_\mathbf{k}^x = 0$ for every $\mathbf{k}$.

In the particular case when the Bloch Hamiltonian exhibits time-reversal symmetry 
[i.e., when the Hamiltonian matrix given in Eq.~(\ref{eqn:H0}) satisfies
$
h_{0\mathbf{k}} = h_{0,-\mathbf{k}}^*
$
or simply
$
d_\mathbf{k}^x = d_{-\mathbf{k}}^x,
$
$
d_\mathbf{k}^y = -d_{-\mathbf{k}}^y
$
and
$
d_\mathbf{k}^z = d_{-\mathbf{k}}^z],
$
a uniformly-condensed Bose gas at the zero-momentum Bloch state 
(i.e., when $\mathbf{k_c = 0}$ and $\theta_\mathbf{0} = \pi/2$) 
has a trivial geometric contribution to the effective-mass tensor of the 
superfluid carriers. 
This is because
$
\dot{\theta}_\mathbf{0}^{\ell} = 0
$
when
$
d_\mathbf{0}^z = 0 = \dot{d}_\mathbf{0}^{z,\ell}.
$ 
See also App. A below.
Thus, when the time-reversal symmetry manifests, a uniformly-condensed Bose 
gas must occur at a finite momentum Bloch state 
(i.e., $\mathbf{k_c \ne 0}$ and $\theta_\mathbf{k_c} = \pi/2$)
in order for a non-trivial geometric contribution to appear. 
Such a situation can only be realized if there exists a degeneracy in the 
single-particle ground state, e.g., in the presence of a spin-orbit coupling 
or in a flat Bloch band. The former possibility has recently been 
addressed in full details~\cite{subasi22}. However, in the latter 
possibility~\cite{you12}, since the bare inverse band-mass tensor 
$(M^{-1})_{\ell\ell'}$ necessarily vanishes in Eq.~(\ref{eqn:Meffg}), 
the low-energy Bogoliubov modes are determined entirely by a particular 
value of the quantum-metric tensor, 
i.e., by $g^{\mathbf{k_c}}_{\ell\ell'}$~\cite{julku21a, julku21b, julku22}.
Example models are given in App. B.

\subsection{Comparison with the Fermi superfluids}
\label{sec:Fs}

Typically the building blocks for the many-body problem in Fermi superfluids 
can be found in the two-body problem, and it turns out the quantum-geometric 
effects are already apparent in this exactly-solvable 
limit~\cite{iskin21, iskin22}. 
For instance, in the presence of time-reversal symmetry and under the condition 
of uniform pairing on all sublattices within a unit cell, the inverse effective-mass 
tensor for the lowest-lying two-body bound-state band has a quantum-geometric 
contribution that is controlled precisely by the quantum-metric tensor of the 
underlying Bloch states. The exact relation is in fact a $\mathbf{k}$-space 
sum over a few terms that can be associated with either the intraband or the 
interband processes, where the band-resolved quantum-metric tensor appears in 
the latter with some additional energy factors~\cite{iskin21, iskin22}.
It turns out the many-body problem is quite similar to the two-body one: 
the inverse effective-mass tensor of the superfluid carriers (i.e., the Cooper 
pairs) also has a quantum-geometric contribution originating from the 
interband processes~\cite{iskin18b, huhtinen22, herzog22}. 
This finding further suggests that all of 
the superfluid properties that depend on the pair mass must have some 
quantum-geometric contribution, including but not limited to the superfluid 
weight/density (i.e., superfluid stiffness) and low-energy collective 
excitations (i.e., Goldstone modes)
~\cite{peotta15, liang17, huhtinen22, herzog22, chan22b, iskin18b, iskin20b}.

\section{Conclusion}
\label{sec:conc}

In summary here we considered a weakly-interacting Bose gas that is described 
by a generic two-band Bose-Hubbard model, and derived its Bogoliubov spectrum 
for the low-energy quasiparticle excitations.
We showed that the interband processes that are induced by the interactions 
give rise to a quantum-geometric contribution, and dress the effective-mass 
tensor of the superfluid carriers. When the BEC occurs uniformly 
within a unit cell (i.e., equal condensate filling on both sublattices), 
we also related the geometric contribution to the quantum-metric tensor of 
the Bloch states, which is nothing but the natural Fubini-Study metric on 
the Bloch sphere. Thus, in the particular case when the bare band-mass 
tensor vanishes (e.g., in a flat Bloch band), the energetic stability of 
the Bogoliubov modes, and therefore the BEC itself, is guaranteed 
by a finite quantum-geometric contribution. This shows that the previous 
results are immune to the presence of non-local 
interactions~\cite{julku21a, julku21b, julku22}. 

Similar to the Bogoliubov spectrum and superfluid weight/density, 
we expect that all of the superfluid properties that depend on the effective 
carrier mass to have analogous quantum-geometric contributions. 
These contributions can be distinguished by their linear dependence on 
the interactions, and are well-worthy of further research and exploration. 
In particular, since our formalism is based on the Bogoliubov approximation, 
our analytical expressions are not valid away from the weakly-interacting 
limit. As the interactions become stronger, we expect the intraband 
contribution coming from the upper Bloch band to affect the effective-mass 
tensor, especially when the interaction energy becomes comparable to the 
total bandwidth of the single-particle spectrum. Note that such a contribution 
plays a negligible role in the weakly-interacting limit, thanks to the band 
gap that is protecting the ground state.

\begin{acknowledgments}
While finalizing this manuscript, Ref.~\cite{julku22} appeared in the 
preprint server, where the speed of Bogoliubov sound is calculated up to 
second-order in the interactions for a flat-band BEC. 
The author acknowledges funding from T{\"U}B{\.I}TAK.
\end{acknowledgments}
\appendix
\subsection*{Appendix A. Benchmark with the extended Bose-Hubbard model}
Here we specifically consider the extended Bose-Hubbard model with only 
nearest-neighbor hopping $t > 0$, onsite repulsion $U$ and nearest-neighbor 
repulsion $V$, and assume time-reversal symmetry. In addition we also 
suppose that the BEC occurs at the zero-momentum Bloch state, and it is 
uniform in a unit cell, i.e., 
$
d_\mathbf{k_c = 0}^z = 0.
$
Under these conditions, there is no geometric contribution to the low-energy 
Bogoliubov modes as discussed in Sec.~\ref{sec:qm}.
As an illustration, we calculate the Bogoliubov spectrum given in 
Eq.~(\ref{eqn:Esq}), and find
\begin{align}
\label{eqn:eBH}
E_{s\mathbf{q}} &= s\sqrt{
\big(\varepsilon_{-,\mathbf{q}} - \varepsilon_{-,\mathbf{0}}\big)
\big(\varepsilon_{-,\mathbf{q}} - \varepsilon_{-,\mathbf{0}} 
+ n_0 U + I_\mathbf{q}\big)}, \\
I_\mathbf{q} &= -2n_0 u_\mathbf{q} v_\mathbf{q} 
\mathrm{Re}\big[U_{AB}(\mathbf{q}) e^{i\varphi_\mathbf{q}}\big],
\end{align}
where we use the chemical potential
$
\mu = \varepsilon_{-,\mathbf{0}} + n_0(\mathcal{U} + \mathcal{V})/2,
$
and $\varphi_{-\mathbf{q}} = -\varphi_\mathbf{q}$. 
Here $\mathcal{U} = U$, $\mathcal{V} = z V$, and $z$ is the coordination 
number, e.g., $z = \{3,4,6\}$ for honeycomb, square and triangular lattices. 
Note that the contribution from the nearest-neighbor interactions can also be 
written as
$
I_\mathbf{q} = n_0 (V/t) [(d_\mathbf{q}^x)^2 + (d_\mathbf{q}^y)^2]/d_\mathbf{q},
$
and Eq.~(\ref{eqn:eBH}) reproduces the usual result when $V = 0$.

As a nontrivial example, let's consider a square lattice with lattice spacing $a$, 
and describe it with a unit cell that contains a two-point basis, i.e., treat it
like a bipartite checkerboard lattice. 
Then its single-particle spectrum is characterized by
$
d_\mathbf{k}^x = -2t[\cos(k_x a) + \cos(k_y a)]
$
and
$
d_\mathbf{k}^y = d_\mathbf{k}^z = d_\mathbf{k}^0 = 0.
$ 
In this case Eq.~(\ref{eqn:eBH}) reproduces the known result~\cite{danshita09} 
in the reduced Brillouin zone, i.e., in a square region bounded 
by $|k_x| + |k_y| = \pi/a$ since the lattice period is doubled in both $x$ 
and $y$ directions, in which $\varphi_\mathbf{k} = \pi$ for every $\mathbf{k}$. 
Note that their condensate filling is defined per lattice site, i.e., $n_0 = 2\nu$.
Thus our projected Hamiltonian and its Bogoliubov theory is quantitatively 
accurate in describing the low-energy physics. 
Furthermore it can be shown (for sufficiently large $V$) that Eq.~(\ref{eqn:eBH}) 
develops a roton minimum at the edges of the reduced Brillouin zone. 
Then by setting, e.g., 
$
(\partial^2 E_{-,\mathbf{q}} / \partial q_x^2)_{\mathbf{q}=(0,\pi/a)} \ge 0
$
at one of the corners, we find that the roton minimum occurs if
$
n_0 (\mathcal{V}-\mathcal{U}) \ge 8t.
$
This condition coincides precisely with the threshold for the dynamical 
instability which signals the superfluid-to-supersolid phase 
transition~\cite{danshita09}.

\subsection*{Appendix B. Example models with non-trivial geometry}

Mielke checkerboard model is one of the simplest two-band lattice models that 
exhibit a flat band in two dimensions. See the supplemental material 
of Ref.~\cite{montambaux18} for a realistic proposal of its implementation using 
optical-lattice potentials. Within our reciprocal-space convention, 
the single-particle spectrum in such a lattice is described by
$
d_\mathbf{k}^0 = 2t \cos(k_x a) \cos(k_y a),
$
$
d_\mathbf{k}^x = 2t \cos(k_x a) + 2t \cos(k_y a)
$
and
$
d_\mathbf{k}^z = 2t \sin(k_x a) \sin(k_y a),
$
leading to a flat lower band 
$
\varepsilon_{-, \mathbf{k}} = -2t
$
and a dispersive upper band
$
\varepsilon_{+, \mathbf{k}} = 2t + 4t \cos(k_xa) \cos(k_ya).
$
Thus the resultant Bloch bands touch at the four corners of the 
Brillouin zone, i.e., at
$
\mathbf{k} = \{(\pm \pi/a, 0), (0, \pm \pi/a)\}.
$ 
Setting $d_\mathbf{k}^z = 0$ shows that there exists a continuous subset 
of flat-band states that favor uniform condensation on sublattices A and B, 
and hence, is expected to minimize the condensation energy. 
Since its flat-band states also admit real representation for every $\mathbf{k}$, 
Eqs.~(\ref{eqn:Meff}) and (\ref{eqn:Meffg}) directly apply to this model.
In addition the model also exhibits time-reversal symmetry, and therefore
a finite $\mathbf{k_c}$ guarantees a non-trivial geometric contribution.

Other two-band models that feature a flat band in two dimensions include 
checkerboard I, II and III lattices~\cite{rhim19}. For instance the latter 
model is described by
$
d_\mathbf{k}^0 = 7t/2 + t\cos(k_x a) + 2t\cos(k_y a),
$
$
d_\mathbf{k}^x = -2t - 2t\cos(k_x a) - t\cos(k_y a) - 2t \cos(k_x a + k_y a),
$
$
d_\mathbf{k}^y = - 2t\sin(k_x a) - t\sin(k_y a) - 2t \sin(k_x a + k_y a)
$
and
$
d_\mathbf{k}^z = 3t/2 - t\cos(k_x a) + 2t\cos(k_y a),
$
and it leads to a flat lower band
$
\varepsilon_{-, \mathbf{k}} = 0
$
that is gapped from the dispersive upper band. The minimum band gap occurs at 
the four corners of the Brillouin zone, i.e., at
$
\mathbf{k} = \{(\pm \pi/a, 0), (0, \pm \pi/a)\}.
$ 
Setting again $d_\mathbf{k}^z = 0$ shows that there exists a continuous subset 
of flat-band states that favor uniform condensation on sublattices A and B, 
and hence, minimize the condensation energy~\cite{julku21b}. 
While this model also exhibits time-reversal symmetry, its flat-band states
do not admit real representation for every $\mathbf{k}$. Thus only 
Eq.~(\ref{eqn:Meff}) applies to this model. See Ref.~\cite{julku21b} for a
detailed analysis of this particular model and its numerical illustration.

\bibliography{refs}

\end{document}